\documentclass[amssymb,prb,aps,twocolumn,amsmath,showpacs]{revtex4}

\usepackage[dvips]{graphicx}
\usepackage{amstext}

\begin{document}

\title{Josephson Junctions with a synthetic antiferromagnetic interlayer}
\author{Mazin A. Khasawneh, W. P. Pratt, Jr., Norman O. Birge}
\email{birge@pa.msu.edu}
\affiliation{Department of Physics and
Astronomy, Michigan State University, East Lansing, Michigan
48824-2320, USA}
\date{\today}

\begin{abstract}
We report measurements of the critical current vs. Co thickness in
Nb/Cu/Co/Ru/Co/Cu/Nb Josephson junctions, where the inner Co/Ru/Co
trilayer is a "synthetic antiferromagnet" with the magnetizations
of the two Co layers coupled antiparallel to each other via the
0.6 nm-thick Ru layer.  Due to the antiparallel magnetization
alignment, the net intrinsic magnetic flux in the junction is
nearly zero, and such junctions exhibit excellent Fraunhofer
patterns in the critical current vs. applied magnetic field, even
with total Co thicknesses as large as 23 nm.  There are no
apparent oscillations in the critical current vs. Co thickness,
consistent with theoretical expectations for this situation. The
critical current of the junctions decays over 4 orders of
magnitude as the total Co thickness increases from 3 to 23 nm.
These junctions may serve as useful templates for future
explorations of spin-triplet superconducting correlations, which
are predicted to occur in superconducting/ferromagnetic hybrid
systems in the presence of certain types of magnetic
inhomogeneity.
\end{abstract}

 \maketitle

Superconducting/ferromagnetic (S/F) hybrid systems have received
much attention in the past decade.\cite{BuzdinReview} When a
conventional spin-singlet Cooper pair crosses the S/F interface,
the two electrons enter different spin bands, hence the pair picks
up a momentum shift proportional to the exchange
energy.\cite{Demler} This physical process leads to a number of
oscillatory phenomena in S/F systems, including oscillations in
the $T_c$ of S/F bilayers and in the critical current of S/F/S
Josephson junctions as a function of F-layer
thickness.\cite{BuzdinReview}  There are proposals to use S/F/S
$\pi$-junctions as components in superconducting circuits or in
various quantum computing schemes.

A relatively recent development is the prediction of a new kind of
spin-triplet pair correlations in S/F induced in conventional S/F
systems by the presence of certain forms of magnetic
inhomogeneity.\cite{Bergeret:01a,Kadigrobov:01}  Unlike
spin-singlet pairs, spin-triplet pairs are not subject to the
exchange field, hence they should propagate long distances in a
ferromagnetic material -- limited only by the temperature or by
spin-flip or spin-orbit scattering. One place to search for
spin-triplet correlations is in thick S/F/S Josephson junctions,
where the spin-singlet supercurrent is exponentially suppressed by
the exchange field.\cite{Keizer:06} Depending on their geometry
and the type of F material, however, thick S/F/S junctions may
contain a large amount of intrinsic magnetic flux, which distorts
the ``Fraunhofer" pattern of the critical current vs. applied
magnetic field and reduces the reliability of critical current
measurements.

We report here measurements of Nb/Cu/Co/Ru/Co/Cu/Nb Josephson
junctions, where the central Co/Ru/Co trilayer is a ``synthetic
antiferromagnet" with the magnetizations of the two Co layers
exchange-coupled antiparallel to each other via the 0.6 nm-thick
Ru layer.\cite{Parkin:90}  The total Co thickness was varied
between 3 and 23 nm -- much thicker than in previous studies of
S/F/S junctions using Co with thicknesses up to 5
nm.\cite{Robinson:06} Over our range of Co thicknesses, the
critical current drops by more than 4 orders of magnitude, while
the critical current vs. applied magnetic field exhibits a nearly
perfect Fraunhofer pattern over the entire range.  We do not
observe any signature of spin-triplet correlations in these
samples, but we suggest that they may serve as a useful platform
for future searches for triplet correlations, perhaps by adding
additional magnetic layers with inhomogeneous magnetization
adjacent to the Nb layers.

\begin{figure}[tbh]
\begin{center}
\includegraphics[width=3.2in]{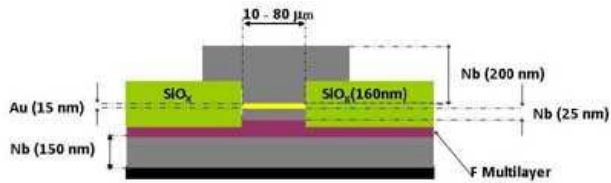}
\end{center}
\caption{(color online). Schematic diagram of S/F/S Josephson
junction cross section, where the "F multilayer" refers to the
Co/Ru/Co trilayer, with or without additional Cu buffer layers
adjacent to the Nb electrodes. Current flow is in the vertical
direction.  The magnetic field is applied in the plane of the
layers, i.e. perpendicular to the current
direction.}\label{Schematic}
\end{figure}

Multilayer samples of the form
Nb(150)/Cu(5)/Co(x)/Ru(0.6)/Co(x)/Cu(5)/Nb(25)/Au(15), with all
thicknesses in nm, were grown by dc triode sputtering in an Ar
pressure of 2.5 mTorr, in a system with base pressure of $2 \times
10^{-8}$Torr.  The thin Cu layers appear to change the growth
characteristics of the Co layers, and result in larger critical
current of the junctions for thick Co layers.  (Results for
samples with and without the Cu layers will be shown below.) The
total Co thickness, $d_{Co} = 2x$, was varied between 3 and 23 nm.
The multilayers were patterned into circular pillars of diameters
10, 20, 40, and 80 $\mu$m using an image reversal
photolithographic process and Ar ion milling.  The milling was
followed immediately by deposition of 160 nm of SiO$_x$, then
lift-off of the photoresist mask.  Finally, top Nb electrodes of
thickness 200 nm were deposited by sputtering.  A schematic
diagram of the sample geometry is shown in Fig. \ref{Schematic}.
All critical current measurements were performed at 4.2 K with the
samples inside a Cryoperm magnetic shield, using a SQUID-based
current comparator method.\cite{EdmundsPratt} Current-voltage
characteristics of all samples followed the standard form for
overdamped Josephson junctions.

\begin{figure}[tbh]
\begin{center}
\includegraphics[width=2.8in]{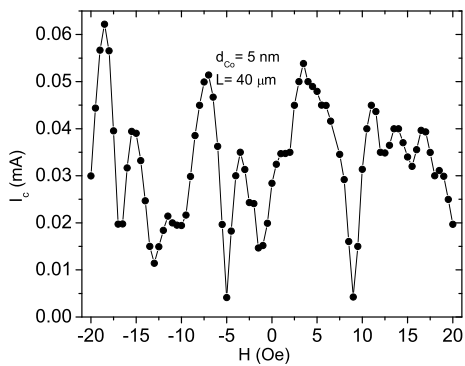}
\end{center}
\caption{Critical current vs. magnetic field applied in the film
plane (perpendicular to the current direction) for a Nb/Co/Nb
circular Josephson junction of diameter 40 $\mu$m and $d_{Co}$= 5
nm.}\label{CoFraun}
\end{figure}

A valuable tool for characterizing the quality of Josephson
junctions is the measurement of critical current vs. magnetic
field applied perpendicular to the current direction. Observation
of a good Fraunhofer pattern for junctions guarantees that the
current flow is uniform across the junction, and that there are no
shorts in the surrounding insulator.  Observation of the
Fraunhofer pattern in S/F/S junctions with strong ferromagnets,
however, can be problematic, due to the intrinsic magnetic flux of
the ferromagnetic domains. For sufficiently thin F layers, the
Fraunhofer patterns can be extremely good.\cite{Sprungmann:09} In
junctions with extremely small lateral dimensions, good Fraunhofer
patterns can be obtained over a larger range of F-layer
thickness.\cite{Robinson:07} But for sufficiently thick F layers,
the Fraunhofer pattern becomes random, with no clear central
maximum.  An example for a circular junction of diameter $40
\mu$m, with a single Co layer 5 nm thick, is shown in Fig.
\ref{CoFraun}. The deep minima in $I_c$ at $H = -5$ and $+8$ Oe
demonstrate that there are no shorts in the oxide surrounding the
junction. The overall pattern, however, is quite random, due to
the randomness of the magnetic domain structure of the Co film.
Similar random Fraunhofer patterns have been seen previously in
S/F/S junctions containing other strong ferromagnetic materials:
Gd\cite{Bourgeois:01} and Ni.\cite{Khaire:09}

\begin{figure}[tbh]
\begin{center}
\includegraphics[width=3.2in]{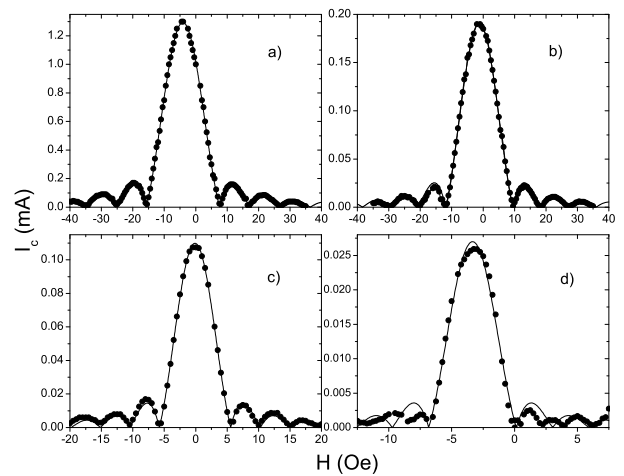}
\end{center}
\caption{Critical current vs. applied magnetic field obtained for
Nb/Cu/Co/Ru/Co/Cu/Nb circular Josephson junctions with different
total thickness of the Co layers: a) 6.1 nm, b) 11nm, c) 18 nm,
and d) 23 nm.  The pillar diameters $w$ are 10, 10, 20, and 40
$\mu$m, respectively.  The solid lines are fits to Eq.
(\ref{Airy}).}\label{FraunSAF}
\end{figure}

\begin{figure}[tbh]
\begin{center}
\includegraphics[width=2.6in]{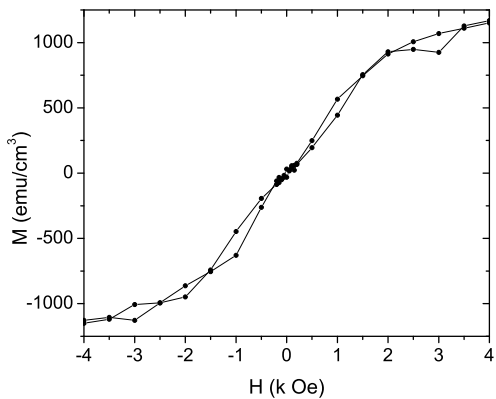}
\end{center}
\caption{Magnetization vs. applied field at $T=10$K for a
Co(4)/Ru(0.6)/Co(4) trilayer grown on 150 nm of Nb.} \label{MvsH}
\end{figure}

Fabrication of Josephson junctions containing the synthetic
antiferromagnetic trilayer, Co(x)/Ru(0.6)/Co(x), circumvents this
problem.  Fig. \ref{FraunSAF} shows Fraunhofer patterns for four
samples with total Co thicknesses varying from 6.1 to 23 nm.  The
first three patterns are nearly perfect, while the last one is
still extremely good.  The maximum field shift of the patterns is
a few Oe, which indicates a very strong antiferromagnetic coupling
between the top and bottom Co layers.  Solid lines are fits to the
theoretical Airy formula for junctions with circular cross
section:
\begin{equation}\label{Airy}
I_c(\Phi)=I_c(0)\frac{2\times
J_{1}(\frac{\pi\Phi}{\Phi_{0}})}{(\frac{\pi\Phi}{\Phi_{0}})},
\end{equation}
where $J_{1}$ is the Bessel function of the first kind, $\Phi_{0}
= h/2e$ is the flux quantum, $\Phi=H_{ext}(2\lambda_L + d)w$ is
the magnetic flux penetrating the junction with $\lambda_L$ the
London penetration depth, $w$ the junction diameter and $d$ the
barrier thickness.  The antiferromagnet coupling of the Co/Ru/Co
trilayer was confirmed independently from measurements of
magnetization ($M$) vs. applied field for Co(4)/Ru/Co(4) trilayers
with varying thicknesses of Ru. The coupling was evident for Ru
thicknesses of 0.6, 0.7, 0.8, and 0.9 nm. Fig. \ref{MvsH} shows
$M$ vs. $H$ for a sample with Ru thickness of 0.6 nm. $M$ doesn't
saturate until $H$ is at least 5 kOe, and there is very little
hysteresis between curves with $H$ increasing and decreasing.

We subjected one Josephson junction sample to a series of large
in-plane magnetic fields, then re-measured $I_c$ vs. $H$ at low
field. The resulting Fraunhofer patterns showed only slight
distortion after applying fields as large as 5 kOe. After applying
10 kOe the central peak in the Fraunhofer pattern split into two
peaks of about half the original magnitude. After warming the
sample to room temperature and cooling back to 4.2 K, an excellent
Fraunhofer pattern was obtained once again.

\begin{figure}[tbh]
\begin{center}
\includegraphics[width=3.2in]{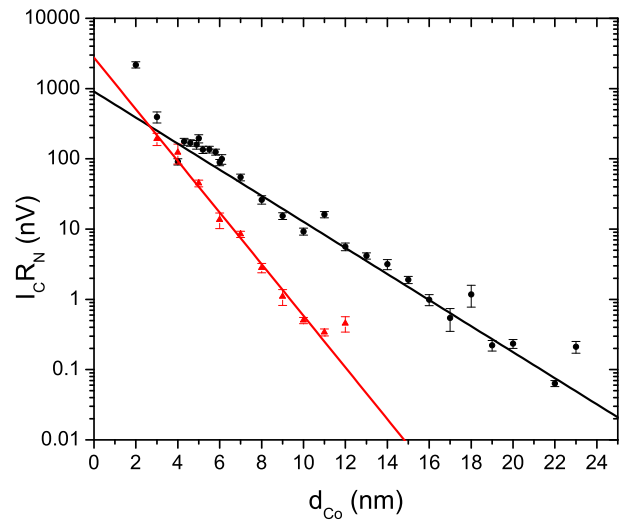}
\end{center}
\caption{(color online). Product of critical current times normal
state resistance vs. total Co thickness for all of our SAF
Josephson junctions. Red points (triangles) are data for samples
without Cu buffer layers, while black points (circles) are data
for samples with Cu buffer layers. Error bars represent the
standard deviation of measurements taken on more than one pillar
on the same substrate, with the minimum uncertainty chosen to be
10\%.  The solid lines are fits to a simple exponential decay,
with decay lengths of $1.18 \pm 0.05$nm and $2.34 \pm 0.08$nm,
respectively.} \label{IcvsdCo}
\end{figure}

The dependence of $I_c$ on total Co thickness ($d_{Co}$) is
summarized in Fig. \ref{IcvsdCo}.  The figure shows two sets of
data: black circles represent the samples fabricated with the Cu
buffer layer, while red triangles represent samples fabricated
earlier without the Cu layer.  In both cases $I_c$ decays
exponentially with $d_{Co}$.  In samples without the Cu, the decay
is faster than in samples with Cu.  We focus first on the larger
data set -- the samples with Cu.  An immediate question is whether
these samples can be $\pi$-junctions; i.e. does $I_c$ oscillate
with $d_{Co}$?  While the data do not display convincing
oscillations, there are a few data points (e.g. for $d_{Co}$=4.0,
18, and 23 nm) that are substantially above or below their
neighbors. To address the question of oscillations, we fabricated
a set of samples in one sputtering run with closely spaced Co
layer thicknesses in the range $4.3-6.1$ nm.  Those samples do not
exhibit any local minima in $I_c$, whereas Robinson \textit{et
al.}\cite{Robinson:06} observed a spacing of 1.0 nm between local
minima for Nb/Co/Nb junctions containing a single Co layer.

Several theoretical papers address the expected behavior of $I_c$
vs. $d_{Co}$ for Josephson junctions containing two magnetic
layers with non-collinear
magnetizations.\cite{BlanterHekking:04,Vedyayev:05,Crouzy:07} We
discuss only the situation relevant to our experiments, where the
two ferromagnetic layers have equal thickness and antiparallel
magnetizations.  In the ballistic limit, such S/F/F/S junctions
are predicted to behave similarly to S/N/S junctions -- with a
slow algebraic decay and no oscillations in $I_c$ -- because the
relative phase shift acquired by the two electrons of a Cooper
pair as they travel through the first F layer is exactly cancelled
by the phase shift they acquire through the second F
layer.\cite{BlanterHekking:04} In the presence of disorder the
critical current decays exponentially with F layer thickness, but
still without any oscillations of the kind associated with S/F/S
junctions.  Our data are consistent with this picture.

To extract quantitative information from our data, we must go a
step further with the theory.  The theoretical works cited above
calculate the exact form of the $I_c$ decay only in certain
limits, e.g. for the pure ballistic case with no elastic
scattering and for the diffusive limit with $E_{ex} \tau \ll
\hbar$, where $\tau$ is the mean free time between collisions.
Josephson junctions with Co, however, fall into an intermediate
limit, where $E_{ex} \tau > \hbar$, but $\Delta \tau \ll \hbar$,
with $\Delta$ the superconducting gap. Although theories in the
intermediate limit do not exist for S/F/F/S junctions, they do
exist for S/F/S junctions,\cite{Bergeret:01b,Kashuba} and predict
exponential decay of $I_c$ with a decay constant equal to the mean
free path in the F material.  (Theories for S/F/S junctions also
predict oscillations, which are not present in the S/F/F/S case
studied here.) The solid lines in Fig. \ref{IcvsdCo} are
least-squares fits of an exponential decay to our two data sets,
with decay lengths $2.34 \pm 0.08$nm for the samples with Cu
buffer layers and $1.18 \pm 0.05$nm for the samples without Cu.
This analysis suggests that the mean free path in the Co grown on
the Cu buffer layer is longer than in the Co grown directly on Nb,
probably due to less strain in the former case.  Confirmation of
that hypothesis would require analysis of the grain structure of
our films by transmission electron microscopy.

It is instructive to compare our results with those of Robinson
\textit{et al.}\cite{Robinson:06,Robinson:07}, who studied S/F/S
junctions made with the strong ferromagnets Co, Ni, Fe, and Py,
all of which are believed to lie in the intermediate limit defined
above. Those workers found that, for Ni and Py, the $I_c$ vs.
$d_F$ data followed an algebraic decay for small $d_F$ and an
exponential decay for larger $d_F$, with the crossover interpreted
as occurring when $d_F$ surpasses the mean free path, $l_e$.  For
Co, the data could be fit with either an algebraic or exponential
decay over the thickness range studied, $0.8-5$nm. As shown in
Fig. \ref{IcvsdCo}, our $I_c$ data decay exponentially over the
entire range of $d_{Co} = 2-23$nm, with the possible exception of
our first data point.  Given the extra scattering in our samples
from the two Co/Ru interfaces,\cite{Ahn:08} it is not surprising
that $l_e < d_{Co}$ over the entire range of Co thicknesses we
measured.  What is more surprising is that, if we were to plot the
data of Robinson \textit{et al.}\cite{Robinson:07} (ignoring the
oscillations) in Fig. \ref{IcvsdCo}, they would lie a factor of
100 higher than our data over the narrow thickness range covered
by both experiments. This suggests that the thin Ru layer severely
suppresses $I_c$, possibly due to spin memory loss at the Co/Ru
interfaces.

The single exponential decay of $I_c$ vs. $d_{Co}$ shown in Fig.
\ref{IcvsdCo} indicates a lack of spin-triplet superconducting
correlations in these samples, which would manifest themselves as
a crossover to a slower decay with increasing $d_{Co}$.  (An
optimist might consider the point at $d_{Co}=23$nm as a hopeful
sign, but a sample with $d_{Co}=24$nm exhibited a very small
supercurrent and no Fraunhofer pattern, hence it was excluded from
the Figure.) There are several possible reasons why we do not
observe the long-range triplet correlations (LRTC). First, there
could be substantial spin memory loss at the Co/Ru interfaces --
an issue we intend to clarify in the near future for our samples
using giant magnetoresistance techniques. Second, the amplitude of
the LRTC generated at the S/F interfaces may be too small to
measure. This could occur either if the domain structure in the Co
films contains mostly domains aligned along a single directions in
space (the LRTC requires non-collinear magnetizations), or if the
LRTC component has random phases at adjacent Co domain walls, and
hence averages to zero over the lateral dimensions of the
samples.\cite{Bergeret-private} The latter situation could be
ameliorated by fabricating samples with smaller lateral
dimensions, while both issues could be addressed by utilizing a
magnetic material with a well-characterized form of magnetic
inhomogeneity, such as the spiral magnetic structure occurring in
materials such as
Ho.\cite{Sosnin:06,VolkovAnishchanka:06,Halasz:09}

In this context, we note that optimizing the \textit{generation}
of the LRTC at the S/F interface may involve a choice of materials
that does not optimize \textit{propagation} of the LRTC through
the subsequent ferromagnetic materials.  It is here where we
believe the Josephson junctions reported in this paper may hold
the most promise.  One can imagine producing samples of the form
S/X/SAF/X/S, where X is a magnetic material chosen to optimize
LRTC generation, while SAF is a suitable synthetic antiferromagnet
with little spin memory loss, either the Co/Ru/Co trilayer studied
here or a weaker SAF such as Co/Cu/Co.\cite{Parkin:91}  Once the
SAF layer becomes sufficiently thick (greater than about 23 nm for
the case of the Co SAF studied here), the singlet supercurrent is
suppressed by over 4 orders of magnitude. Generation of the LRTC
at the S/X interfaces would then be manifested as a long-range
spin-triplet supercurrent that persists out to SAF thicknesses far
beyond what has been measured here.

We are grateful to S. Bergeret, Y. Blanter, D. van Harlingen, A.
Volkov, and K. Efetov for helpful discussions, and to B. Bi, T.
Khaire, R. Loloee, and Y. Wang for technical assistance. This work
was supported by the Department of Energy under grant
DE-FG02-06ER46341.

\end{document}